\documentclass{iaus}
\usepackage{graphicx}
\usepackage{natbib}

\title[Radio and mid-IR observations of MYSO candidates] 
{The RMS Survey: Radio and mid-IR observations of candidate massive YSOs}

\author[J.~C.~Mottram and J.S. Urquhart et al.:]   
{J.~C.~Mottram, J.~S. Urquhart, M.~G.~Hoare, S.~L.~Lumsden, \and R.~D.~Oudmaijer
   }

\affiliation{School of Physics and Astronomy, University of Leeds, Leeds, LS2~9JT, UK \break email: jcm@ast.leeds.ac.uk}
\pubyear{2006}
\date{August 9-11 2006 Visegrad, Hungary}
\jname{Interaction of Stars with their Environment III}
\editors{Editor: V. Toth}
\begin{document}

\maketitle

\vspace*{-0.25cm}
\begin{abstract}

The Red MSX Source (RMS) survey is a multi-wavelength programme of follow-up observations designed to distinguish between genuine massive young stellar objects (MYSOs) and other embedded or dusty objects, such as ultra compact (UC) HII regions, evolved stars and planetary nebulae (PNe), from a sample of $\sim$ 2000 MYSOs candidates. These were identified by comparing the colours of objects from the MSX and 2MASS point source catalogues to those of known MYSOs, in order to develop colour selection criteria which have been used to produced the RMS sample of MYSOs candidates. Our ultimate aim is to produce a large unbiased sample of MYSOs ($\sim$~500) with complementary multi-wavelength data with which to study their properties. Here we report the results of 826 radio continuum and 346 mid-IR imaging observations carried out using the Australia Telescope Compact Array and TIMMI2 on the ESO 3.6~m telescope respectively. These observations are aimed at identifying and removing contaminating sources.

\end{abstract}

\vspace*{-0.75cm}
\section*{Introduction}

\noindent During their relatively short lives massive O and early B type stars can have an enormous impact on their local environments. They inject huge amounts of energy into the interstellar medium (ISM) in the form of radiation and via molecular outflows, powerful stellar winds and supernova explosions. The radiation emitted is predominantly in the form of UV-photons which ionise their natal molecular cloud, leading to the formation of an ultra compact (UC) HII region. Though initially deeply embedded, these ionised regions expand, breaking free of their natal cloud and eventually evolving into the more classical HII regions. The radiation also heats the surrounding material, evaporating ice mantles from the surfaces of dust particles, which alters the local chemistry. Throughout their lives massive stars process a huge amount of material and are responsible for the production of most of the heavy elements, which are returned to the ISM upon the stars' death, ejected and/or created in the subsequent supernova explosion. Through these processes, massive stars play an important role in the enrichment of the ISM.

Given that massive stars have such a profound impact, not only on their local environment, but also on a Galactic scale, understanding the environmental conditions and processes involved in their birth and the earliest stages of their evolution are of fundamental importance. However, their formation and the early stages of their evolution are still poorly understood, since they take place within dense cores of dust and molecular gas, which are opaque to traditional optical and UV probes. Additionally, massive stars are very rare, and are thus generally located much farther away than regions of low-mass star formation. They are known to form predominantly in clusters, making it difficult to attribute derived quantities to individual sources. This is compounded by the fact that massive young stellar objects (MYSOs) evolve much more rapidly than in the case of low-mass stars, reducing the number available for study still further.

MYSOs are mid-infrared bright with luminosities of 10$^4$--10$^6$ L$_\odot$ (\citealt{wynn-williams1982}). Many have been associated with massive bipolar molecular outflows (e.g., \citealt{wu2004}), and  therefore accretion is still likely to be ongoing, although nuclear fusion has almost certainly begun. The large accretion rates are thought to quench the ionising radiation, and in so doing impede the formation of a UCHII region (\citealt{forster2000}). MYSOs are also known to possess ionising stellar winds, but these winds are weak thermal radio sources ($\sim$~1~mJy at 1 kpc; \citealt{hoare2002}). These objects can therefore be roughly parameterised as mid-IR bright and relatively radio continuum quiet. \citet{lumsden2002} compared colours of objects from the MSX and 2MASS point source catalogues (\citealt{egan2003} and \citealt{skruskie2006} respectively) to those of known MYSOs, in order to develop colour selection criteria, which were used to produced an unbiased sample of approximately 2000 candidate MYSOs. 

There are several other types of embedded, or dust enshrouded objects, that have similar colours to MYSOs which contaminate our sample, such as UCHII regions, evolved stars and planetary nebulae (PNe). However, these contaminating sources can be identified by incorporating information obtained from other wavelengths. We are currently involved in a multi-wavelength programme of follow-up observations known as the Red MSX Source (RMS) survey (\citealt{hoare2005}). These observations are designed to distinguish between genuine MYSOs and other embedded or dusty objects, and also to provide further data with which to characterise the general properties of MYSOs. These include
high resolution radio cm continuum observations to identify UCHII regions and PNe (\citealt{urquhart2006}), molecular line observations to obtain kinematic distances (e.g., \citealt{busfield2006}) and bolometric luminosities, mid-IR imaging to identify genuine point sources, evaluate the issues of clustering and to obtain accurate positions, and near-IR spectroscopy (e.g., \citealt{clarke2006}) to distinguish between MYSOs and evolved stars. Here we present the results of our radio continuum and mid-IR observations.

\vspace*{-0.5cm}
\section*{Radio observations}

\noindent Our radio continuum observations were made using the Australia Telescope Compact Array (ATCA). These consisted of five snapshots of each source with an integration time of two minutes, resulting in an image rms of $\sim0.3$~mJy, a field of view of $\sim6$--10\arcmin\ with a angular resolution of a few arcseconds. Radio emission was detected within a 12$^{\prime\prime}$ radius of 25\% of the RMS sample observed (199 out of 826 RMS sources observed; see \citealt{urquhart2006} for details). These sources are therefore not genuine MYSOs but are most likely embedded UCHII regions. Although this effectively eliminates a quarter of the RMS sources observed, the remaining 75\% (627 sources) towards which no radio emission was observed are still MYSO candidates. In total we found 211 radio sources to be associated with 199 RMS sources, with multiple radio sources being associated with a single RMS source in a number of  cases;  these multiple radio source matches could imply real clustering of UCHII regions.

In Fig. 1 we present plots of the two dimensional distribution of all RMS sources observed with the ATCA, and the RMS sources associated with radio emission (upper and lower panels respectively). This figure shows the distribution of the RMS sources to be correlated with the Galactic plane, however, the RMS sources that are associated with radio emission display an even tighter correlation with the Galactic plane. The distributions look similar and have a similar angular scaleheight $\sim0.6$--0.7\degr, in excellent agreement with the value found by \cite{woodchurchwell1989} who found it to be $\sim0.6$\degr\ from IRAS identification of $\sim$~1600 potential UCHII regions. From an analysis of the radio fluxes of all RMS-radio sources detected at both frequencies (excluding those found to have complex morphologies) we find the average spectral index to be $-$0.16, consistent with emission from thermal radio sources (e.g., UCHII regions). Additionally, we find the distribution of spectral indices is slightly skewed in the positive direction, possibly indicating that a significant number of optically thick sources have been detected.

Therefore the majority of the 199 RMS sources found to be associated with radio emission are expected to be UCHII regions, since our original colour selection criteria remove all but heavily reddened PNe. The morphologies of these sources, which are consistent with other studies of UCHII regions, relatively flat spectral indices, and their scaleheight certainly support their identification as UCHII regions. Once combined with the results of our VLA observations (Urquhart et al. in prep.) we will have an unbiased sample of $\sim$~400-500 UCHII regions, the vast majority of which were previously unknown.

\begin{figure*}
\begin{center}
\includegraphics[width=0.98\textwidth,trim= 20 50 0 50 ]{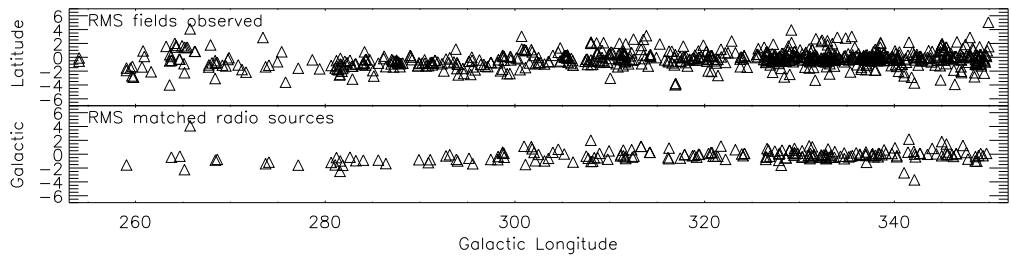}

\caption{The upper and lower panels on the left we presented plots of the two dimensional distribution of the all RMS sources observed with the ATCA and the RMS sources associated with radio emission respectively.}

\end{center}
\end{figure*}

\vspace*{-0.5cm}
\section*{Mid-IR observations}

\noindent The observed mid-IR size of an object has important implications when considering its nature. From simulations in the literature (e.g., \citealt{churchwell1990}; their Fig. 2(a)) the computed full-width half-maximum of an MYSO at 12~$\mu$m is $\sim10^{16}$~cm ($\sim0.003$~pc). At a distance of just 1~kpc an MYSO will therefore only cover 0.6\arcsec\ on the sky, and so should be unresolved in our 1\arcsec\ resolution observations. The primary contaminant of our sample, as mentioned above, is UCHII regions. Resonantly scattered Lyman $\alpha$ photons are the main source of heating for the dust, leading to the radio and mid-IR morphologies being very similar (\citealt{hoare1991}). Many UCHII regions will be resolved at 1\arcsec\ resolution, while those that are unresolved can be identified by their compact radio emission. We therefore expect MYSOs to be unresolved point-sources while UCHII regions and PN will appear predominantly as extended objects (see Fig.~2). Evolved stars will also be unresolved and so require identification by other means, e.g. near-IR spectroscopy or lack of CO emission. Where 10~$\mu$m imaging improves over radio continuum observations alone is that at mid-IR wavelengths, MYSOs near UCHII regions will appear as point sources near extended emission (e.g. see Fig.~2) while radio observations would only detect the UCHII region. This prevents any bias against MYSOs located near UCHII regions.

\begin{figure*}
\begin{center}
\includegraphics[width=0.49\textwidth]{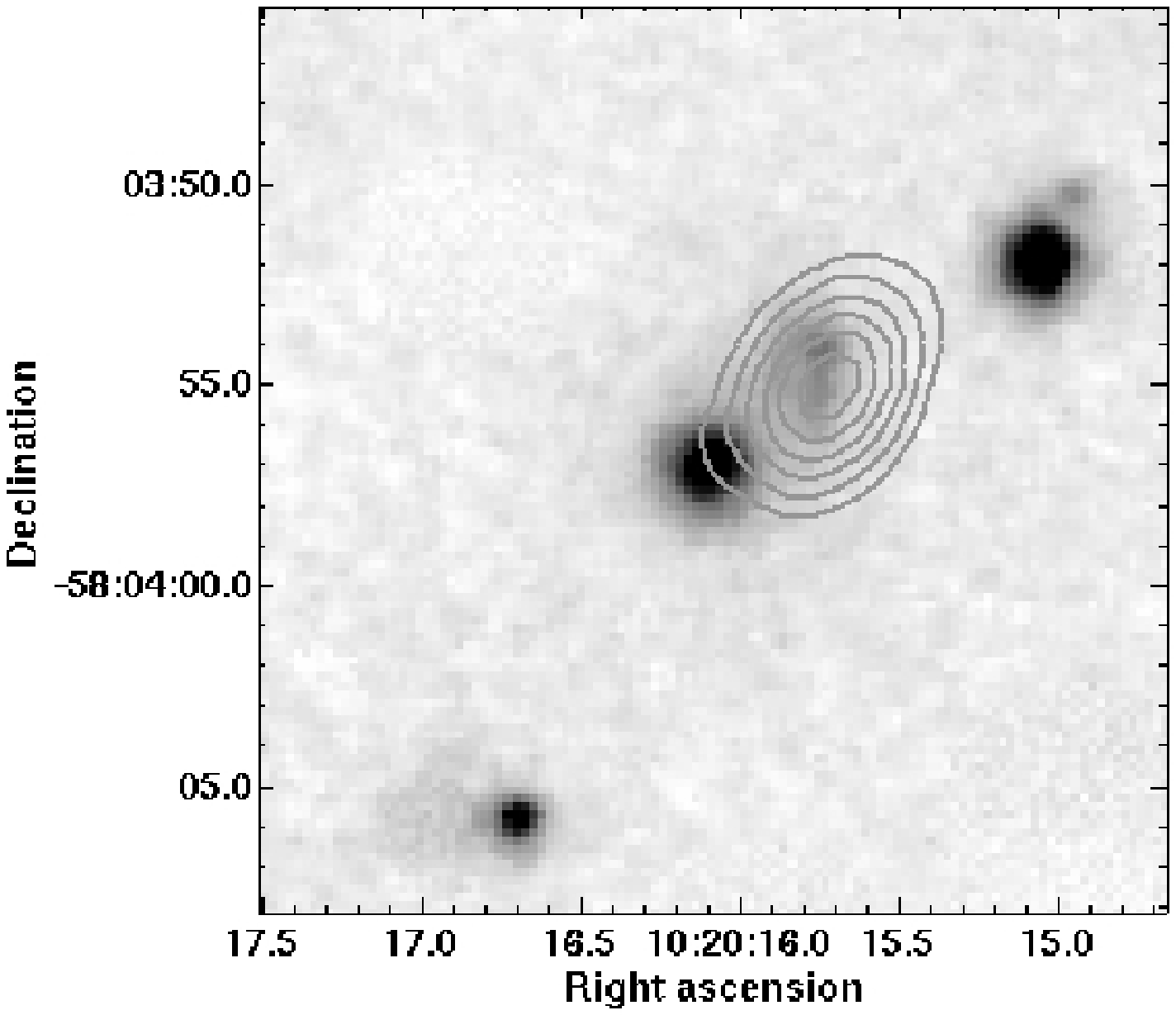}
\includegraphics[width=0.50\textwidth]{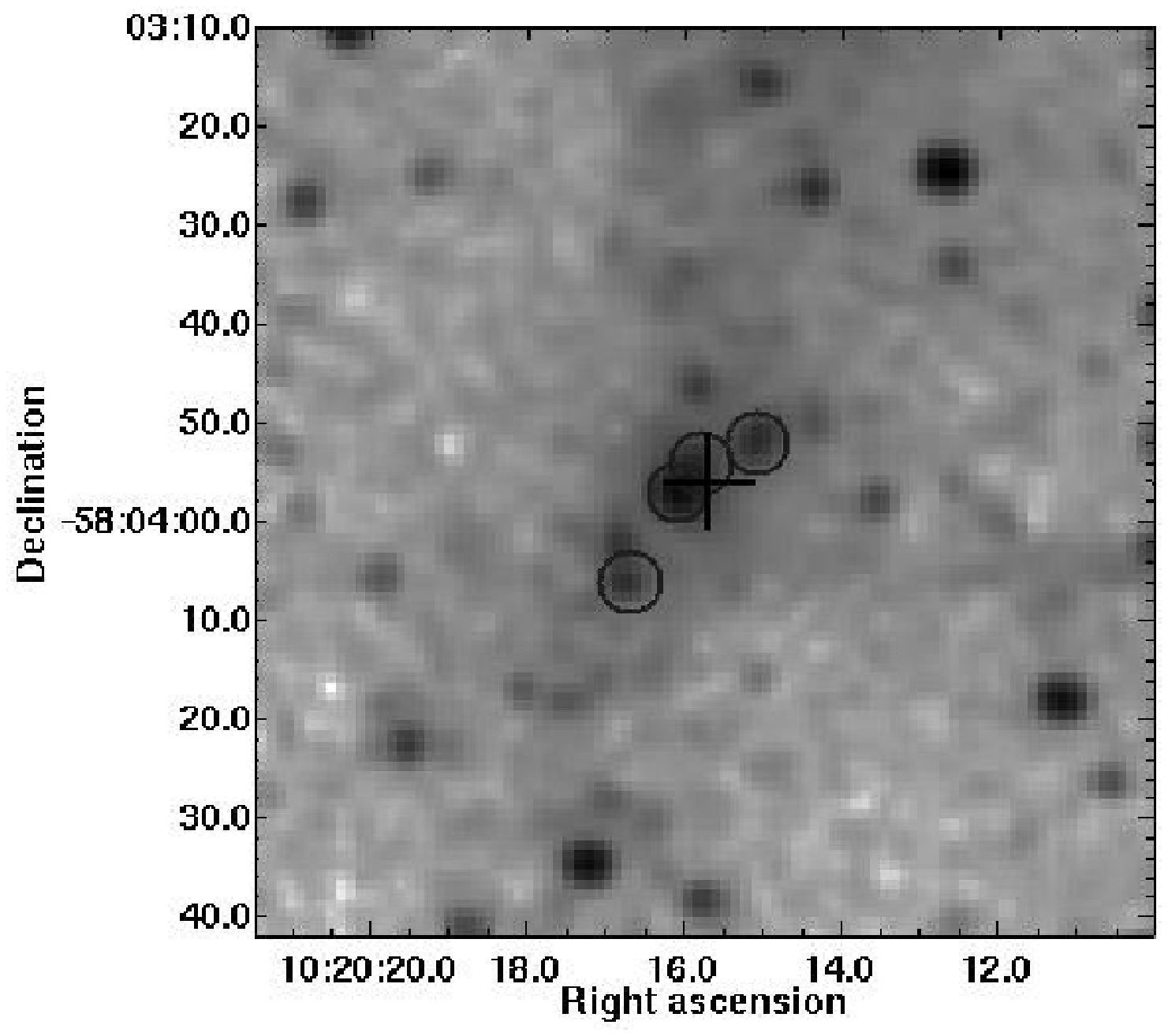}
\caption{Left: A 10~$\mu$m image of G284.0155-00.8579 with a 99$\%$ scale cut to emphasise the extended emission near the central point, with 6~cm radio contours from our ATCA data overlaid. Right: $K$ band 2MASS image of the same region with MSX (+) and TIMMI2 ($\circ$) detections indicated.}
\end{center}
\end{figure*}

Our mid-IR 10~$\mu$m imaging observations were taken using TIMMI2 on the ESO 3.6~m telescope at La Silla, Chile with a field of view of 64$^{\prime\prime} \times 48^{\prime\prime}$ and a resolution of $\sim1^{\prime\prime}$. We observed 346 targets, with the preliminary results that roughly 12\% are non-detections at a 3$\sigma$ upper limit of $\sim$~0.05~Jy. These are most likely extended HII regions where the mid-IR surface brightness is too weak to be detected in our observations. Approximately 50\% of our targets are isolated point sources, $\sim15\%$ are isolated extended regions, $\sim8\%$ contain multiple point sources, $\sim3\%$ contain multiple extended sources and $\sim12\%$ contain complexes of both point and extended sources. Overall, $\sim30\%$ of the targets observed contain sources of extended emission, which is similar to the findings of our radio observations that $\sim25\%$ of RMS sources are associated with radio emission, consistent with the targets being UCHII regions (see previous section). 

Our photometry, with 3$\sigma$ accuracies of the order 0.05~Jy, will allow us to apportion larger beam fluxes (e.g. MSX) where multiple sources are detected, and to eliminate from our sample those point sources whose luminosity in larger beam observations is dominated by nearby, more diffuse emission. Our astrometric accuracy was of order 2\arcsec\, which will provide improved information for our near- and mid-IR spectroscopy observations.

\vspace*{-0.5cm}
\section*{Summary}

\noindent The RMS survey will provide a large, unbiased sample of order 500 MYSOs from a colour-selected list of $\sim2000$ candidates using a series of follow-up observations. These include mid-IR imaging, radio continuum and molecular line observations, and near-IR spectroscopy. Mid-IR imaging allows us to apportion the flux from larger beam observations between multiple targets if detected, and provides improved astrometric information for spectroscopy. Our preliminary results on the number of extended sources at mid-IR wavelengths are consistent with the findings of our radio observations that $\sim25\%$ of our candidate targets are UCHII regions.

\vspace*{-0.5cm}
\bibliography{hung}

\bibliographystyle{aa}



\end{document}